\documentclass[letter,twocolumn]{jpsj3}
\usepackage{txfonts}
\usepackage{color}
\usepackage{bm}

\title
{
Almost Perfect Frustration in the Dimer Magnet Ba$_2$CoSi$_2$O$_6$Cl$_2$
}

\author
{
Hidekazu \textsc{Tanaka}$^{1}$\thanks{E-mail address: tanaka@lee.phys.titech.ac.jp}, Nobuyuki \textsc{Kurita}$^{1}$, Makiko \textsc{Okada}$^{1}$, Eiji \textsc{Kunihiro}$^{1}$, Yutaka \textsc{Shirata}$^{1}$, Kotaro \textsc{Fujii}$^{2}$, Hidehiro \textsc{Uekusa}$^{2}$, Akira \textsc{Matsuo}$^3$, Koichi \textsc{Kindo}$^3$, Hiroyuki \textsc{Nojiri}$^4$
}

\inst
{
$^{1}$Department of Physics, Tokyo Institute of Technology, Meguro-ku, Tokyo 152-8551\\
$^{2}$Department of Chemistry, Tokyo Institute of Technology, Meguro-ku, Tokyo 152-8551\\
$^{3}$Institute for Solid State Physics, University of Tokyo, Kashiwa, Chiba 277-8581\\
$^{4}$Institute for Materials Research, Tohoku University, Aoba-ku, Sendai 980-8577
}

\recdate
{
\hspace{7cm}
}

\abst{
We determined the crystal structure of Ba$_2$CoSi$_2$O$_6$Cl$_2$, which was synthesized in this work, and investigated its quantum magnetic properties using single crystals. This compound should be described as a two-dimensionally coupled spin-1/2 XY-like spin dimer system. Ba$_2$CoSi$_2$O$_6$Cl$_2$ exhibits a stepwise magnetization process with a plateau at half of the saturation magnetization, irrespective of the field direction, although all the Co$^{2+}$ sites are equivalent. This indicates that spin triplets are localized owing to the almost perfect frustration of interdimer exchange interactions. Thus, the spin states for the zero and 1/2 magnetization-plateau states are almost exactly given by the simple product of singlet dimers and the alternate product of singlet and triplet dimers, respectively.
}

\begin{document}
\maketitle

Coupled spin dimer magnets provide a stage to embody the quantum physics of interacting lattice bosons~\cite{Rice}. These magnets often exhibit a gapped singlet ground state. In an external magnetic field exceeding the energy gap, the $S_z\,{=}\,{+}1$ component of the spin triplet is created on the dimer. The $S_z\,{=}\,{+}1$ component, called a magnon, acts as a boson on the dimer lattice. The number of magnons can be tuned via the external magnetic field corresponding to the chemical potential. Magnons move to neighboring dimers and interact with one another owing to the transverse and longitudinal components of the interdimer exchange interactions, respectively. When the hopping term is dominant, magnons can undergo Bose-Einstein condensation (BEC)~\cite{Nikuni,Matsumoto,Giamarchi,Oosawa1,Oosawa2,Rueegg,Jaime,Waki,Zapf,Stone,Yamada}, while when the hopping term is suppressed so that the repulsive interaction due to the antiferromagnetic interdimer interaction is dominant, magnons can crystallize to form a regular array~\cite{Kageyama,Miyahara1,Onizuka,Momoi,Kodama,Miyahara2,Takigawa,Matsuda}. Coupled spin dimer magnets also have the potential to realize an exotic supersolid phase in a magnetic field~\cite{Chen,Albuquerque,Murakami,Yamamoto}. 

Many spin dimer magnets have been reported to exhibit the BEC of magnons~\cite{Oosawa1,Oosawa2,Rueegg,Jaime,Waki,Zapf,Stone,Yamada}, while the experimental realization of magnon crystallization has been limited.
It is known that an orthogonal spin dimer system, which was first discussed theoretically by Shastry and Sutherland~\cite{Shastry} and is realized in SrCu$_2$(BO$_3$)$_2$~\cite{Kageyama}, undergoes the successive crystallization of magnons with fractional magnetization plateaus~\cite{Miyahara1,Momoi,Onizuka,Kodama,Miyahara2,Takigawa,Matsuda}. In this letter, we report the crystal structure of the spin dimer magnet Ba$_2$CoSi$_2$O$_6$Cl$_2$ and show that this compound exhibits magnon crystallization in a high magnetic field owing to the strong frustration of interdimer exchange interactions, which is different from the orthogonal dimer model.

Ba$_2$CoSi$_2$O$_6$Cl$_2$ single crystals were synthesized as a by-product of the process used for the crystal growth of Ba$_3$CoSb$_2$O$_9$, which is an $S\,{=}\,1/2$ triangular-lattice Heisenberg antiferromagnet~\cite{Shirata,Susuki,Zhou_PRL2012}. Because the crystal structure of Ba$_2$CoSi$_2$O$_6$Cl$_2$ has not been reported to date, we performed a structural analysis of it at room temperature. Details of the crystal growth and the analysis of the crystal structure are described in the Supplemental Materials~\cite{Supplement}. 

\begin{figure}[htb]
\begin{center}
\includegraphics[width=6.5cm, clip]{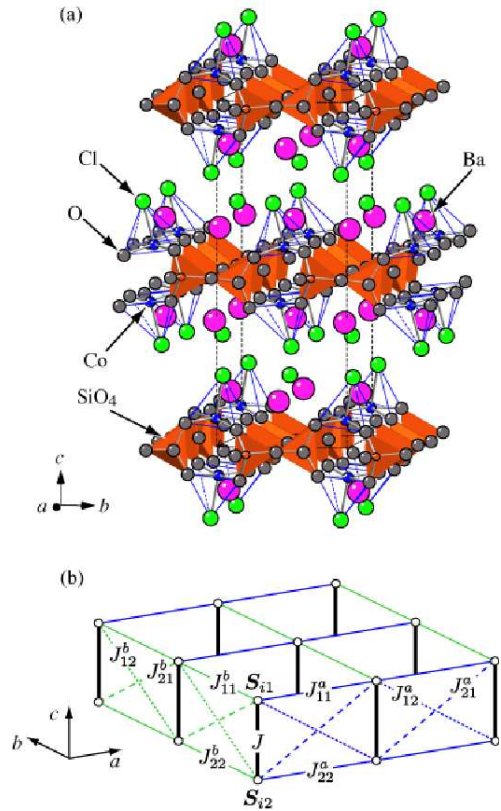}
\end{center}
\caption{(Color online) (a) Perspective view of the crystal structure of Ba$_2$CoSi$_2$O$_6$Cl$_2$. Dotted lines denote the chemical unit cell. A magnetic Co$^{2+}$ ion is located approximately at the center of the base of the CoO$_4$Cl pyramid, where the pyramids are linked via orange-colored SiO$_4$ tetrahedra in the $ab$ plane. (b) Model of the exchange network for Ba$_2$CoSi$_2$O$_6$Cl$_2$. Thick solid lines represent the intradimer exchange interaction $J$, and thin solid, dashed and dotted lines represent the interdimer exchange interactions $J^a_{{\alpha}{\beta}}$ and $J^b_{{\alpha}{\beta}}\ ({\alpha},{\beta}=1, 2)$.}
 \label{fig:cryst}
\end{figure}

The structure of Ba$_2$CoSi$_2$O$_6$Cl$_2$ is monoclinic $P2_1/c$ with cell dimensions of $a\,{=}\,7.1382$\,$\rm{\AA}$, $b\,{=}\,7.1217$\,$\rm{\AA}$, $c\,{=}\,18.6752$\,$\rm{\AA}$ and ${\beta}\,{=}\,91.417^{\circ}$. A perspective view of the crystal structure is illustrated in Fig.~\ref{fig:cryst}(a). The crystal structure consists of CoO$_4$Cl pyramids with a Cl$^-$ ion at the apex. Magnetic Co$^{2+}$ is located approximately at the center of the base composed of O$^{2-}$, which is approximately parallel to the $ab$ plane. Two CoO$_4$Cl pyramids form a chemical dimer with their bases facing each other. The CoO$_4$Cl pyramids are linked via SiO$_4$ tetrahedra in the $ab$ plane, as shown in Fig.~S2(b) in the Supplemental Materials. Note that the atomic linkage in the $ab$ plane is similar to that of BaCuSi$_2$O$_6$~\cite{Finger,Sparta,Sasago} (Fig.~S2(c)), which is a model substance undergoing magnon BEC~\cite{Jaime}. Because dispersion measurements~\cite{Sasago} and extensive density-functional calculations of the exchange interactions~\cite{Mazurenko} revealed that the interdimer exchange interactions are dominant in the dimer layer parallel to the $ab$ plane and negligible between dimer layers, we consider that similar interdimer exchange interactions are also realized in Ba$_2$CoSi$_2$O$_6$Cl$_2$. Thus, it is considered that Ba$_2$CoSi$_2$O$_6$Cl$_2$ closely approximates a 2D coupled spin dimer system with the exchange network shown in Fig.~\ref{fig:cryst}(b). This exchange model is closely related to those discussed theoretically~\cite{Chen,Albuquerque,Murakami,Lin,Richter,Derzhko}.

The magnetic property of Co$^{2+}$ in an octahedral environment is determined by the lowest orbital triplet $^4T_1$~\cite{Abragam,Lines}. This orbital triplet splits into six Kramers doublets owing to spin-orbit coupling and the low-symmetric crystal field. When the temperature $T$ is much lower than the magnitude of the spin-orbit coupling constant ${\lambda}\,{=}\,{-}\,178$ cm$^{-1}$, i.e., $T\,{\ll}\, |{\lambda}|/k_{\rm B}\,{\simeq}\,250$\,K, the magnetic property is determined by the lowest Kramers doublet, and the effective magnetic moment of Co$^{2+}$ is represented by the fictitious spin-$1/2$ operator~\cite{Abragam,Lines}. When the octahedral environment exhibits tetragonal symmetry, the effective exchange interaction between fictitious spins ${\bm s}_i$ is described by the spin-$1/2$ XXZ model~\cite{Lines}
\begin{eqnarray}
{\cal H}_{\rm ex}=\sum_{<i,j>} \left[J^{\perp}\left\{s_i^xs_j^x+s_i^ys_j^y\right\}+J^{\parallel}s_i^zs_j^z\right].
\label{eq:int}
\end{eqnarray}
The crystal field in a pyramidal environment, as observed in Ba$_2$CoSi$_2$O$_6$Cl$_2$, can be regarded as that in an octahedral environment with a large tetragonal elongation. In such a case, $J^{\parallel}/J^{\perp}$ is much smaller than unity, so that the effective exchange interaction becomes strongly XY like~\cite{Lines}. The $g$-factors for magnetic fields parallel and perpendicular to the elongated axis of the octahedron are given as $g^{\parallel}\,{=}\,4\,{-}\,5$ and $g^{\perp}\simeq 2$, respectively~\cite{Abragam}. Because the inclination of the magnetic principal $z$ axis from the crystallographic $c^*$ axis is approximately $10^{\circ}$ in Ba$_2$CoSi$_2$O$_6$Cl$_2$, we assume that the condition of $H\,{\parallel}\,c^*$ corresponds to that of $H\,{\parallel}\,z$. The energy levels of the isolated dimer are given by $-(2J^{\perp}\,{+}\,J^{\parallel})/4$, $J^{\parallel}/4$ and $(2J^{\perp}\,{-}\,J^{\parallel})/4$, and the corresponding eigenstates are $|0,0\rangle$, $|1,{\pm}1\rangle$ and $|1,0\rangle$, respectively.

The magnetic susceptibility of Ba$_2$CoSi$_2$O$_6$Cl$_2$ single crystal was measured in the temperature range of $1.8\,{-}\,300$ K using a SQUID magnetometer (Quantum Design MPMS XL). The specific heat of Ba$_2$CoSi$_2$O$_6$Cl$_2$ was measured  in the temperature range of $0.4\,{-}\,300$ K using a physical property measurement system (Quantum Design PPMS) by the relaxation method. High-field magnetization measurement in a magnetic field of up to 70 T was performed at 4.2 and 1.3 K using an induction method with a multilayer pulse magnet at the Institute for Solid State Physics, University of Tokyo. 
High-frequency ESR measurements in pulsed magnetic fields of up to 30 T with fixed frequencies ranging from 190 to 450 GHz were performed at 4.2 K at the Institute for Materials Research, Tohoku University. Ba$_2$CoSi$_2$O$_6$Cl$_2$ crystal is easily cleaved parallel to the $ab$ plane. Magnetic fields were applied parallel to the $ab$ plane and the $c^*$ axis, which is perpendicular to the $ab$ plane.

\begin{figure}[t] 
\begin{center}
\includegraphics[width=7.5cm, clip]{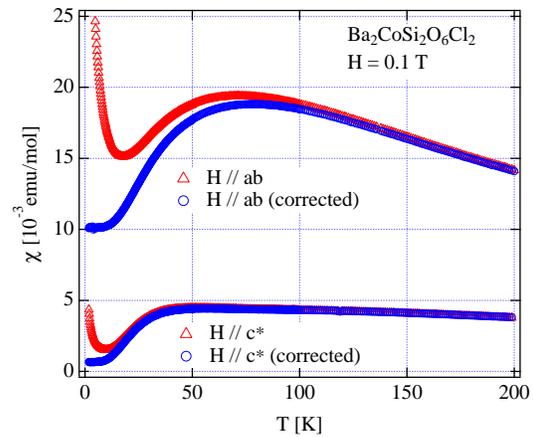}
\end{center}
\caption{(Color online) Magnetic susceptibilities in Ba$_2$CoSi$_2$O$_6$Cl$_2$ measured at $H\,{=}\,0.1$ T for $H\,{\parallel}\,c^*$ and $H\,{\parallel}\,ab$. Red and blue symbols are raw data and the data corrected for the Curie term. 
} 
\label{fig:sus}
\end{figure} 

Figure~\ref{fig:sus} shows the magnetic susceptibilities in Ba$_2$CoSi$_2$O$_6$Cl$_2$ measured for magnetic fields $H$ parallel to the $c^*$ axis and $ab$ plane. The magnetic susceptibility is isotropic in the $ab$ plane. This is because there are domains in which the $a$ and $b$ directions are interchanged, which was observed by the present X-ray diffraction measurement. The Curie term below 10 K is ascribed to unpaired spins produced by lattice defects because the magnitude of the Curie term depends strongly on the field direction. The blue symbols in Fig.~\ref{fig:sus} show the magnetic susceptibilities corrected for the Curie term. For both field directions, the magnetic susceptibilities exhibit a rounded maxima at $50-75$ K and decrease with decreasing temperature. No anomaly indicative of magnetic ordering was observed. The magnetic susceptibility for $H\,{\parallel}\,c^*$ becomes almost zero at $T\,{=}\,0$, while for $H\,{\parallel}\,ab$, the magnetic susceptibility has a finite value even at $T\,{=}\,0$, which is attributed to the large Van Vleck paramagnetism of Co$^{2+}$ in an octahedral (pyramidal) environment. The Van Vleck paramagnetic susceptibility of ${\chi}_{\rm VV}\,{=}\,1.01\,{\times}\,10^{-2}$ emu/mol for $H\,{\parallel}\,ab$ coincides with that obtained from the magnetization curve shown below. The magnetic susceptibility indicates a gapped singlet ground state. It is natural to assume that two Co$^{2+}$ spins located on the bases of neighboring CoO$_4$Cl pyramids are coupled to form an antiferromagnetic dimer, which gives rise to the singlet ground state.

\begin{figure}[htb]
\begin{center}
\includegraphics[width=8.0cm, clip]{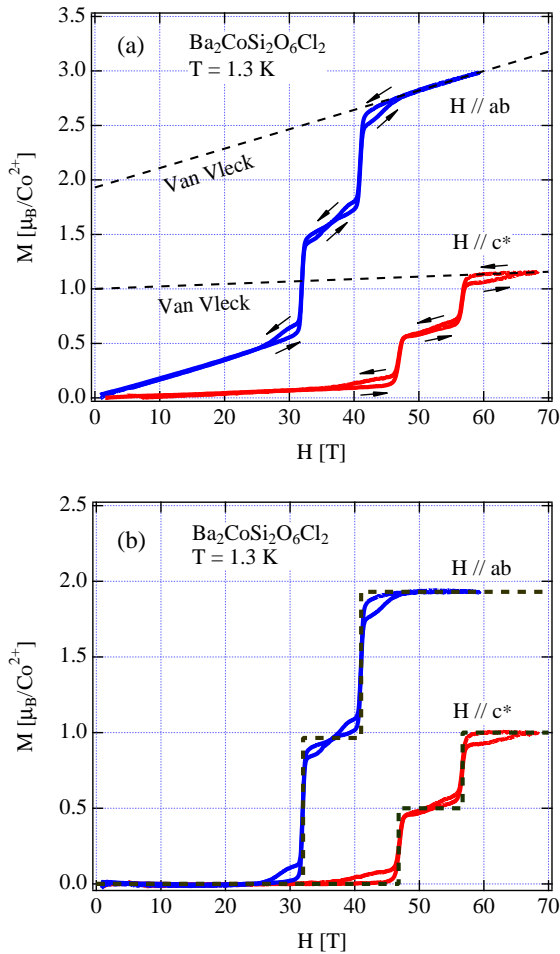}
\end{center}
\caption{(Color online) (a) Raw magnetization curves measured at 1.3 K for $H\,{\parallel}\,c^*$ and $H\,{\parallel}\,ab$ using a piece of single crystal. Arrows indicate the route of the magnetization process upon sweeping the field up and down. (b) Magnetization curves corrected for Van Vleck paramagnetism. Dashed lines are theoretical magnetization curves calculated using eq.\,(\ref{eq:Hc_Hs}) with the parameters given in the text.} 
\label{fig:MH}
\end{figure}

Figure \ref{fig:MH}(a) shows the raw magnetization curves for Ba$_2$CoSi$_2$O$_6$Cl$_2$ single crystal measured at 1.3 K for $H\,{\parallel}\,c^*$ and $H\,{\parallel}\,ab$. The entire magnetization process was observed up to a magnetic field of 70 T. The saturation of the Co$^{2+}$ spin occurs at $H_{\rm s}\,{\simeq}\,57$ and 41 T for $H\,{\parallel}\,c^*$ and $H\,{\parallel}\,ab$, respectively. The absolute value of the magnetization for $H\,{\parallel}\,ab$ was calibrated with the $g$-factor determined as $g_{ab}\,{=}\,3.86$ from the present electron spin resonance (ESR) measurements performed at 4.2 K, as shown below. The $g$ factor for $H\,{\parallel}\,c^*$ was evaluated as $g_{c^*}\,{=}\,2.0\,{\pm}\,0.1$ from the average of the magnetization data. These highly anisotropic $g$-factors of Co$^{2+}$ arise from the pyramidal environment, as previously mentioned~\cite{Abragam,Lines}.
 
The same magnetization slopes below 30 T and above $H_{\rm s}$ for $H\,{\parallel}\,ab$ arise from the large temperature-independent Van Vleck paramagnetism of Co$^{2+}$ in the octahedral environment~\cite{Lines}. From the magnetization slope, the Van Vleck paramagnetic susceptibility for $H\,{\parallel}\,ab$ was evaluated as ${\chi}_{\rm VV}\,{=}\,9.94\,{\times}\,10^{-3}$ emu/mol, which coincides with ${\chi}_{\rm VV}\,{=}\,1.01\,{\times}\,10^{-2}$ emu/mol obtained from the magnetic susceptibility. 
For both field directions, stepwise magnetization processes with a plateau at half of the saturation magnetization $M_{\rm s}$ are observed. The transition fields are $H^{\parallel}_{\rm c}\,{=}\,46.8$ T and $H^{\parallel}_{\rm s}\,{=}\,56.7$ T for $H\,{\parallel}\,c^*$ and $H^{\perp}_{\rm c}\,{=}\,32.0$ T and $H^{\perp}_{\rm s}\,{=}\,41.0$ T for $H\,{\parallel}\,ab$. Note that the magnetization process shows hysteresis, the route of which is indicated by arrows. The magnetization curves measured upon sweeping the field up and down are symmetric with respect the midpoint of the $M_{\rm s}/2$ plateau.

Figure\ \ref{fig:MH}(b) shows the magnetization curves corrected for Van Vleck paramagnetism. It can be clearly seen that the magnetization processes for both field directions are stepwise with a plateau at $M_{\rm s}/2$. The singlet ground state is stabilized in a wide field range below $H_{\rm c}$. The field ranges of the magnetization slopes below and above $M_{\rm s}/2$ plateau are negligible. 

\begin{figure}[t]
\vspace{0.8mm}
\begin{center}
\includegraphics[width=7.5cm, clip]{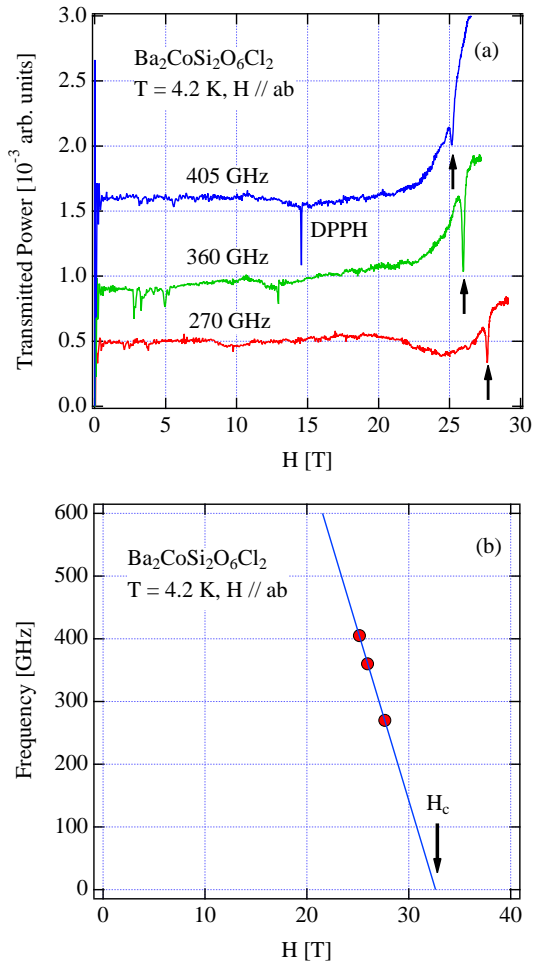}
\end{center}
\caption{(Color online) (a) ESR absorption spectra in Ba$_2$CoSi$_2$O$_6$Cl$_2$ measured at $T\,{=}\,4.2$ K for $H\,{\parallel}\,ab$. Arrows are resonance fields for singlet-triplet transitions. (b) Frequency vs field diagram of ESR mode measured for $H\,{\parallel}\,ab$. The solid line is a linear fit for the singlet-triplet transitions with $g_{ab}\,{=}\,3.86$.}
 \label{fig:ESR}
\end{figure}

We performed ESR measurements on Ba$_2$CoSi$_2$O$_6$Cl$_2$ at low temperatures to obtain the $g$-factor. Figure \ref{fig:ESR}(a) shows ESR absorption spectra for $H\,{\parallel}\,ab$. We were not able to observe electron paramagnetic resonance (EPR) because of the large magnetic anisotropy, but we observed singlet-triplet transitions at the magnetic fields indicated by arrows. The resonance data are summarized in the frequency vs field diagram shown in Fig.\ \ref{fig:ESR}(b). The frequency of the singlet-triplet transition decreases linearly with increasing magnetic field and softens at the critical field $H_{\rm c}^{\perp}$. Assuming that the resonance condition of the singlet-triplet transition near $H_{\rm c}^{\perp}$ is described as ${\hbar}{\omega}\,{=}\,g_{ab}{\mu}_{\rm B}(H_{\rm c}^{\perp}\,{-}\,H)$, we obtain the $g$-factor for $H\,{\parallel}\,ab$ as $g_{ab}\,{=}\,3.86$. We were not able to observe the singlet-triplet transition for $H\,{\parallel}\,c^*$ because the highest magnetic field for which ESR measurement was possible was limited to 30 T.

\begin{figure}[t]
\begin{center}
\includegraphics[width=7.5cm, clip]{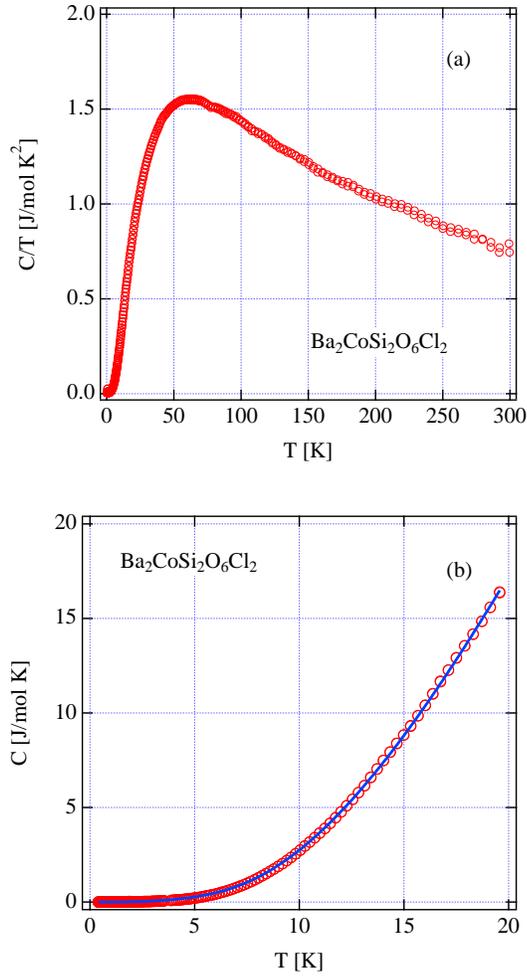}
\end{center}
\caption{(Color online) (a) Specific heat divided by the temperature measured at zero magnetic field. (b) Low-temperature specific heat. The solid line is a fit based on the isolated dimer model with $J^{\perp}/k_{\rm B}\,{=}\,105$ K and $J^{\parallel}/J^{\perp}\,{=}\,0.149$ (fixed).}
\label{fig:heat} 
\end{figure}

There are two possibilities leading to this stepwise magnetization process. The first possibility is that there are two types of almost isolated dimers with equal populations. Magnetization plateaus owing to this mechanism are realized in NH$_4$CuCl$_3$~\cite{Shiramura}. 
In the room-temperature structure, all the Co$^{2+}$ sites are equivalent. Thus, if there are two types of magnetic dimers, a structural phase transition should occur below room temperature. Figure \ref{fig:heat}(a) shows the specific heat divided by the temperature measured at zero magnetic field below 300 K. No anomaly indicative of a phase transition was observed down to 0.5 K. In BaCuSi$_2$O$_6$, which has a similar crystal structure to that of Ba$_2$CoSi$_2$O$_6$Cl$_2$, the interdimer exchange interaction gives rise to a wide magnetization slope with a field range of 26 T~\cite{Jaime}. Thus, the magnitude of the interdimer exchange interaction in Ba$_2$CoSi$_2$O$_6$Cl$_2$ is expected to be of the same order as that in BaCuSi$_2$O$_6$. From these facts, the first possibility is ruled out.

The second possibility is that the interdimer exchange interactions are perfectly frustrated, which corresponds to the case $J^{a,b}_{11}\,{+}\,J^{a,b}_{22}\,{=}\,J^{a,b}_{12}\,{+}\,J^{a,b}_{21}$ within the exchange network shown in Fig.\,\ref{fig:cryst}(b). In this case, the magnon cannot move onto the neighboring dimer, because the hopping amplitude is proportional to $J^{a,b}_{11}\,{+}\,J^{a,b}_{22}\,{-}\,J^{a,b}_{12}\,{-}\,J^{a,b}_{21}$. Hence, the ground state is determined by the balance between the Zeeman energy and the repulsive interaction between magnons that is given by the total of the interdimer exchange interactions, $J^{a,b}_{11}\,{+}\,J^{a,b}_{22}\,{+}\,J^{a,b}_{12}\,{+}\,J^{a,b}_{21}$. 
Consequently, a stepwise magnetization process with the $M_{\rm s}/2$ plateau occurs, as observed in Ba$_2$CoSi$_2$O$_6$Cl$_2$. If the frustration between interdimer interactions is imperfect, i. e., $J^{a,b}_{11}\,{+}\,J^{a,b}_{22}\,{\neq}\,J^{a,b}_{12}\,{+}\,J^{a,b}_{21}$, then the magnetization slopes appear between $M\,{=}\,0$ and the $M_{\rm s}/2$ plateaus and between the $M\,{=}\,M_{\rm s}/2$ plateau and the saturation.\cite{Tachiki2}

When the interdimer exchange interactions are perfectly frustrated, the critical field $H_{\rm c}$ and the saturation field $H_{\rm s}$ are given as
\begin{eqnarray}
 \left.
    \begin{array}{l}
H^{\parallel}_{\rm c}=\displaystyle\frac{J^{\perp}+J^{\parallel}}{2g^{\parallel}{\mu}_{\rm B}},\hspace{2mm}H^{\parallel}_{\rm s}=H^{\parallel}_{\rm c}+\frac{{\alpha}^{\parallel}}{g^{\parallel}{\mu}_{\rm B}},\vspace{2mm}\\
H^{\perp}_{\rm c}=\sqrt{\displaystyle\frac{J^{\perp}(J^{\perp}+J^{\parallel})}{2(g^{\perp}{\mu}_{\rm B})^2}},\vspace{2mm}\\
H^{\perp}_{\rm s}=H^{\perp}_{\rm c}\sqrt{\left(1+\displaystyle\frac{{\xi}{\alpha}^{\perp}}{J^{\perp}}\right)\left(1+\displaystyle\frac{2{\xi}{\alpha}^{\perp}}{J^{\perp}+J^{\parallel}}\right)},\vspace{2mm}\\
{\alpha}^{{\perp},{\parallel}}=\displaystyle\frac{1}{2}\hspace{-0.5mm}\sum_{{\sigma}=a,b}\hspace{-0.7mm}\left(J^{\sigma\,{\perp},{\parallel}}_{11}+J^{\sigma\,{\perp},{\parallel}}_{22}+J^{\sigma\,{\perp},{\parallel}}_{12}+J^{\sigma\,{\perp},{\parallel}}_{21}\right),\vspace{2mm}\\
{\xi}=(H^{\perp}_{\rm s})^2/\left\{(H^{\perp}_{\rm s})^2+(J^{\perp}-J^{\parallel})^2/16\right\}\hspace{-0.5mm}.
    \end{array}
\hspace{-1mm}  \right\}
\label{eq:Hc_Hs}
\end{eqnarray}
The dashed lines in Fig.\ \ref{fig:MH}(b) are magnetization curves calculated using eq.~(\ref{eq:Hc_Hs}) with $J^{\perp}/k_{\rm B}\,{=}\,110$\,K, $J^{\parallel}/J^{\perp}\,{=}\,0.149$, ${\alpha}^{\perp}/k_{\rm B}\,{=}\,23.7$\,K, ${\alpha}^{\parallel}/k_{\rm B}\,{=}\,13.3$\,K, $g_{ab}\,{=}\,3.86$ and $g_{c^*}\,{=}\,2.0$. The field range of the 1/2-plateau is given by ${\alpha}^{{\parallel},{\perp}}/g{\mu}_{\rm B}$. Note that the anisotropy of the interdimer exchange interaction, ${\alpha}^{\parallel}/{\alpha}^{\perp}\,{=}\,0.56$, is different from that of the intradimer exchange interaction, $J^{\parallel}/J^{\perp}\,{=}\,0.15$. In the first approximation, these anisotropies should be the same. We discuss the mechanism leading to the difference in the Supplemental Materials.\cite{Supplement} 

When the hopping of the magnon is absent, the specific heat at low temperatures, where the interaction between thermally excited magnons is negligible, is approximately expressed by that for an isolated dimer. Thus, we describe the low-temperature specific heat as $C\,{=}\,C_{\rm mag}\,{+}\,C_{\rm latt}$, where $C_{\rm mag}$ is the specific heat of the isolated dimer with energy levels of $-(2J^{\perp}\,{+}\,J^{\parallel})/4$, $J^{\parallel}/4$ and $(2J^{\perp}\,{-}\,J^{\parallel})/4$, and $C_{\rm latt}$ is the lattice contribution given by $C_{\rm latt}\,{=}\,a_1T^3\,{+}\,a_2T^5$. The solid line in Fig.~\ref{fig:heat}(b) is the fit based on this model with $J^{\perp}/k_{\rm B}\,{=}\,105$ K and $C_{\rm latt}\,{=}\,0.00213\,T^3\,{-}\,(1.14{\times}10^{-6})\,T^5$ J/mol\,K, where $J^{\parallel}/J^{\perp}$ is fixed to be 0.149. The magnitude of $J^{\perp}/k_{\rm B}$ obtained by this analysis is consistent with $J^{\perp}/k_{\rm B}\,{=}\,110$\,K obtained from the analysis of the magnetization process.

In conclusion, we have presented the results of structural analysis and magnetization, ESR and thermodynamic measurements on Ba$_2$CoSi$_2$O$_6$Cl$_2$ synthesized in this work. This compound should be described as a 2D coupled spin dimer system with XY-like exchange interactions. Ba$_2$CoSi$_2$O$_6$Cl$_2$ exhibits a stepwise magnetization process with an $M_{\rm s}/2$ plateau, irrespective of the magnetic field directions. This finding, together with the absence of a structural phase transition, shows that the frustration for the interdimer exchange interactions is almost perfect. The $M_{\rm s}/2$ plateau state is almost exactly given by the alternate product of singlet and triplet dimers, which corresponds to a Wigner crystal of magnons.

We thank M. Nakamura, Y. Murakami and S. Nishimoto for useful discussions and comments. This work was supported by a Grant-in-Aid for Scientific Research (A) from Japan Society for the Promotion of Science, and by the Global COE Program ``Nanoscience and Quantum Physics'' at Tokyo Tech. funded by the Ministry of Education, Culture, Sports, Science and Technology of Japan.

\section*{Supplemental Materials}
\renewcommand{\thefigure}{S\arabic{figure}}
\renewcommand{\thetable}{S\Roman{table}}

\section*{Crystal growth of Ba$_2$CoSi$_2$O$_6$Cl$_2$}
\begin{figure}[tbh]
\begin{center}
\vspace{1mm}
\includegraphics[width=4.5cm, clip]{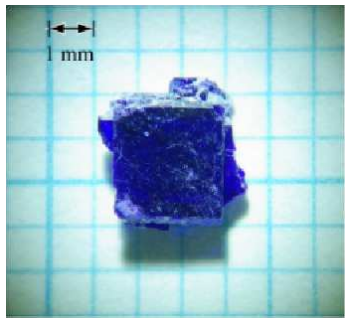}
\end{center}
\caption{Photograph of Ba$_2$CoSi$_2$O$_6$Cl$_2$ single crystal. The wide plane is the crystallographic $ab$ plane. }
\label{fig:S1}
\end{figure}

Ba$_2$CoSi$_2$O$_6$Cl$_2$ single crystals were grown as a by-product of the process used for the single-crystal growth of Ba$_3$CoSb$_2$O$_9$, which is well described as an $S\,{=}\,1/2$ triangular-lattice Heisenberg antiferromagnet~\cite{Shirata}. We used a flux method with BaCl$_2$ as the flux using a quartz tube as a crucible. Using the solid-state reaction, we first prepared Ba$_3$CoSb$_2$O$_9$ powder. A mixture of Ba$_3$CoSb$_2$O$_9$ and BaCl$_2$ in a molar ratio of $1\,{:}\,10$ was sealed in an evacuated quartz tube. The temperature at the center of the furnace was lowered from 1100 to 800$^{\circ}$C over 10 days then rapidly lowered to room temperature. Bluish purple plate-shaped single crystals with a typical size of $3\,{\times}\,3{\times}\,0.5$ mm$^3$ were obtained (Fig.\,\ref{fig:S1}). We first considered that the crystals obtained would be the dimer compound Ba$_2$CoCl$_6$\,\cite{Assoud,Meyer}. However, from X-ray diffraction and the X-ray fluorescence analysis, the crystals were found to be Ba$_2$CoSi$_2$O$_6$Cl$_2$. The silicon was provided from the quartz tube. The wide plane of the crystals was found to be the crystallographic $ab$ plane by X-ray diffraction. It was also found that much larger Ba$_2$CoSi$_2$O$_6$Cl$_2$ single crystals can be grown from a mixture of Ba$_3$CoNb$_2$O$_9$ and BaCl$_2$ in a molar ratio of $1\,{:}\,10$ using a quartz tube.

\section*{Crystal structure}
We performed a structural analysis of Ba$_2$CoSi$_2$O$_6$Cl$_2$ at room temperature using a RIGAKU R-AXIS RAPID three-circle diffractometer equipped with an imaging plate area detector. Monochromatic Mo-K$\alpha$ radiation (${\lambda}\,{=}\,0.71075$\,\rm{\AA}) was used as the X-ray source. Data integration and global-cell refinements were performed using data in the range of $3.06^{\circ}\,{<}\,{\theta}\,{<}\,27.44^{\circ}$, and absorption correction based on face indexing and integration on a Gaussian grid was also performed. The total number of reflections observed was 9097, among which 2163 reflections were found to be independent and 2055 reflections were determined to satisfy the criterion $I > 2{\sigma}(I)$. Structural parameters were refined by the full-matrix least-squares method using SHELXL-97 software. The final $R$ indices obtained were $R\,{=}\,0.0616$ and $wR\,{=}\,0.1787$. The crystal data are listed in Table \ref{table:1}. The chemical formula was found to be Ba$_2$CoSi$_2$O$_6$Cl$_2$ from the results of structural and chemical analyses.

\begin{table}
\vspace{3mm}
\caption{Crystal data for Ba$_2$CoSi$_2$O$_6$Cl$_2$.}
\label{table:1}
\begin{center}
\begin{tabular}{cccc}
\hline
& Chemical formula & Ba$_2$CoSi$_2$O$_6$Cl$_2$ &  \\
& Space group & $P2_1/c$ &  \\
& $a$ ($\rm{\AA}$) & 7.1382(5) &  \\
& $b$ ($\rm{\AA}$) & 7.1217(4) &  \\
& $c$ ($\rm{\AA}$) & 18.6752(10) &  \\
& $\beta$ (deg) &91.417(2) &  \\
& $V$ ($\rm{\AA}^3$) & 949.08(10) &  \\
& $Z$ & 4 &  \\
& $R;\ wR$ &  0.0616;\ 0.1787 & \\
\hline
\end{tabular}
\end{center}
\end{table}

\begin{table}
\caption{Fractional atomic coordinates (${\times}\,10^4$) and equivalent isotropic displacement parameters ($\rm{\AA}^2{\times}\,10^3$) for Ba$_2$CoSi$_2$O$_6$Cl$_2$.}
\label{table:2}
\begin{tabular}{rrrrr}
\hline
Atom  &  $x$\hspace{7mm}    & $y$\hspace{7mm}   & $z$\hspace{5mm}  & $U_{\rm eq}$\hspace{1mm}   \\ \hline
Ba(1)  &  $-$\,5098(1)  &  50(1)  &  1423(1)  &  21(1)  \\
Ba(2)  &  221(1)  &  $-$\,5225(1)  &  1353(1)  &  18(1)  \\
Co(1)  &  $-$\,54(2)  &  	$-$\,115(2)  &  827(1)  &  16(1)  \\
Si(1)  &  	$-$\,2797(4)  &  2771(4)  &  $-$\,7(2)  &  13(1)  \\
Si(2)  &  	$-$\,2775(4)  &  $-$\,2780(4)  &  115(2)  &  13(1)  \\
Cl(1)  &  	$-$\,4554(6)  &  $-$\,481(7)  &  3092(2)  &  40(1)  \\
Cl(2)	  &  624(5)  &  	593(5)  &  2090(2)  &  29(1)  \\
O(1)  &  1934(12)  &  $-$\,2081(12)  &  	746(4)  &  19(2)  \\
O(2)  &  1870(14)  &  1823(13)  &  567(5)  &  25(2)  \\
O(3)  &  $-$\,2068(13)  &  1851(13)	  &  733(5)  &  	22(2)  \\
O(4)  &  $-$\,2034(12)  &  $-$\,2117(12)  &  892(4)  &  20(2)  \\
O(5)  &  $-$\,2320(30)  &  5007(13)	  &  45(6)  &  57(4)  \\
O(6)  &  $-$\,5017(14)  &  2370(30)  &  $-$\,63(7)  &  69(5)  \\
\hline
\end{tabular}
\end{table}

\begin{figure*}[tbh]
\begin{center}
\vspace{2mm}
\includegraphics[width=14.5cm, clip]{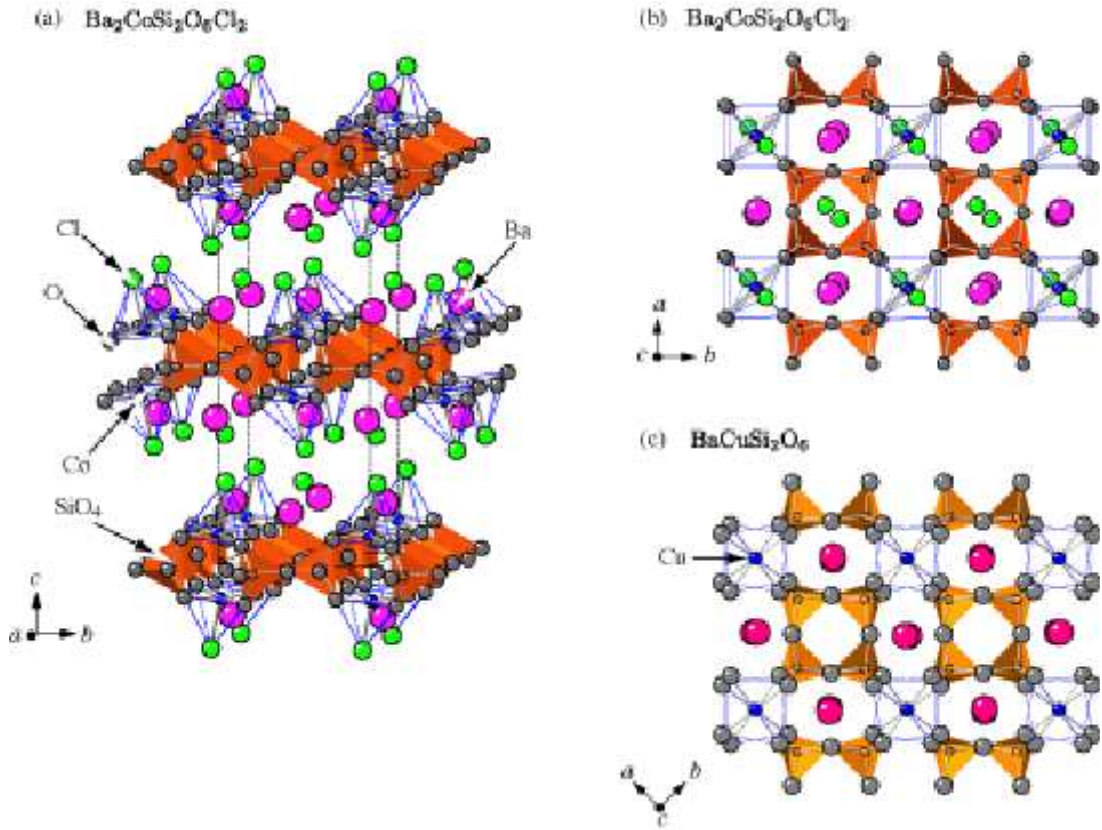}
\end{center}
\caption{(a) Perspective view of the crystal structure of Ba$_2$CoSi$_2$O$_6$Cl$_2$. Dotted lines denote the chemical unit cell. A magnetic Co$^{2+}$ ion is located approximately at the center of the base of the CoO$_4$Cl pyramid, where the pyramids are linked via orange-colored SiO$_4$ tetrahedra. (b) Crystal structure of Ba$_2$CoSi$_2$O$_6$Cl$_2$ viewed from the $c$ axis. The CoO$_4$Cl pyramids are linked via SiO$_4$ tetrahedra in the $ab$ plane. This structure is closely related to that for (c) BaCuSi$_2$O$_6$, in which CuO$_4$ plaquettes are linked via SiO$_4$ tetrahedra in the $ab$ plane\,\cite{Sparta}.}
\label{fig:S2}
\end{figure*}

\begin{figure*}[htb]
\begin{center}
\includegraphics[width=16cm, clip]{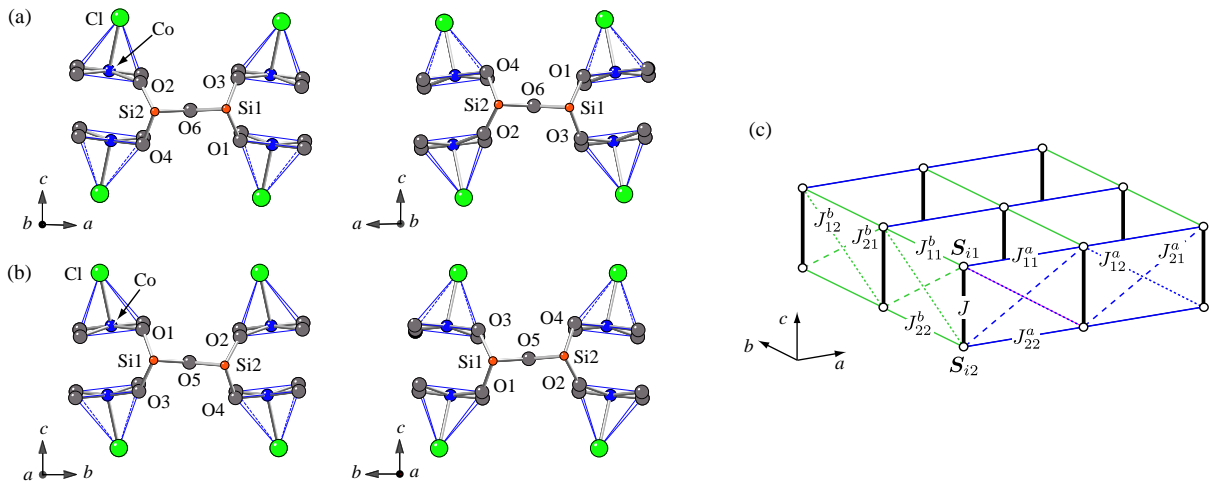}
\end{center}
\caption{Atomic configuration between neighboring dimers (a) along the $a$ direction and (b) along the $b$ direction. (c) A model of exchange network for Ba$_2$CoSi$_2$O$_6$Cl$_2$. Thick solid lines represent the intradimer exchange interaction $J$, and thin solid, dashed and dotted lines represent the interdimer exchange interactions $J_{{\alpha}{\beta}}^a$ and $J_{{\alpha}{\beta}}^b$ (${\alpha}, {\beta}\,{=}\,1, 2$).}
\vspace{-8mm}
\label{fig:S3}
\end{figure*}

The structure of Ba$_2$CoSi$_2$O$_6$Cl$_2$ is monoclinic $P2_1/c$ with cell dimensions of $a\,{=}\,7.1382$\,$\rm{\AA}$, $b\,{=}\,7.1217$\,$\rm{\AA}$, $c\,{=}\,18.6752$\,$\rm{\AA}$, ${\beta}\,{=}\,91.417^{\circ}$ and $Z$=4. Its atomic coordinates and equivalent isotropic displacement parameters  are shown in Table \ref{table:2}. 
A perspective view of the crystal structure and that viewed along the $c$ axis are illustrated in Figs.\,\ref{fig:S2}(a) and (b), respectively. The crystal structure consists of CoO$_4$Cl pyramids, in which Cl$^-$ occupies the apex and magnetic Co$^{2+}$ is located approximately at the center of the base. Two CoO$_4$Cl pyramids form a chemical dimer with their bases facing each other. The CoO$_4$Cl pyramids are linked via SiO$_4$ tetrahedra in the $ab$ plane, as shown in Fig.\,\ref{fig:S2}(b). The atomic linkage in the $ab$ plane is similar to that of BaCuSi$_2$O$_6$\,\cite{Sparta} shown in Fig.\,\ref{fig:S2}(c). In BaCuSi$_2$O$_6$, CuO$_4$ plaquettes are linked via SiO$_4$ tetrahedra in the $ab$ plane and two neighboring Cu$^{2+}$ ions in the $c$ direction form a magnetic dimer\,\cite{Sasago}. Two neighboring CuO$_4$ plaquettes in BaCuSi$_2$O$_6$ are rotated in opposite directions around the $c$ direction, while in Ba$_2$CoSi$_2$O$_6$Cl$_2$, such rotation of the CoO$_4$Cl pyramids is absent.\\
\ \par

\section*{Interdimer exchange interactions}
It is natural to assume that two neighboring Co$^{2+}$ ions along the $c$ direction form a magnetic dimer. Figure\,\ref{fig:S3} shows the atomic configuration between neighboring dimers along the $a$ and $b$ directions and a model of the interdimer exchange interaction in the $ab$ plane. The interdimer exchange interactions $J_{11}^a$ ($J_{11}^b$) and $J_{22}^a$ ($J_{22}^b$) that connects two Co$^{2+}$ spins on the top and bottom, respectively, are equivalent because the midpoint between the neighboring Co$^{2+}$ dimers is the crystallographic inversion center. Thus, the conditions $J_{11}^a=J_{22}^a$ and $J_{11}^b=J_{22}^b$ are satisfied in Ba$_2$CoSi$_2$O$_6$Cl$_2$. 

The contribution of the paths Co$-$O2$-$Si2$-$O6$-$Si1 $-$O3$-$Co and Co$-$O1$-$Si1$-$O6$-$Si2$-$O4$-$Co to the interdimer exchange interaction $J_{11}^a$ closely approximates that of the paths Co$-$O2$-$Si2$-$O6$-$Si1$-$O1$-$Co and Co$-$O1$-$Si1$-$O6$-$Si2$-$O2$-$Co to $J_{12}^a$, because the configurations of $p_{\sigma}$ orbitals for O6 and O3 (O4) in the former two paths are close to those for O6 and O1 (O2) in the latter two paths. The contribution of the paths Co$-$O2$-$O6$-$O3$-$Co and Co$-$O1$-$O6$-$O4$-$Co to $J_{11}^a$ should be close to that of the paths Co$-$O2$-$O6$-$O1$-$Co and Co$-$O1$-$O6$-$O2$-$Co to $J_{12}^a$, because the configurations of $p_{\pi}$ orbitals for O6 and O3 (O4) in the former two paths are similar to those for O6 and O1 (O2) in the latter two paths. This argument is applicable to the $J_{21}^a$ interaction and to those along the $b$ direction. As shown in Figs.\,\ref{fig:S3}(a) and (b), the atomic positions along the $a$ and $b$ directions are close to each other. Therefore, we infer that the conditions $J_{11}^a=J_{22}^a \simeq J_{12}^a \simeq J_{21}^a \simeq J_{11}^b=J_{22}^b\simeq J_{12}^b \simeq J_{21}^b$ are realized in Ba$_2$CoSi$_2$O$_6$Cl$_2$. This situation is different from that in BaCuSi$_2$O$_6$, in which $J_{11}^{a,b}$ and $J_{22}^{a,b}$ are considerably different from $J_{12}^{a,b}$ and $J_{21}^{a,b}$, respectively\,\cite{Mazurenko}. In BaCuSi$_2$O$_6$, the oxygen atom that is common to the two SiO$_4$ tetrahedra connecting CuO$_4$ plaquettes shifts along the $c$ direction, causing the linkage of the two SiO$_4$ tetrahedra to be buckled\,\cite{Sparta}.

As shown in Fig.\,\ref{fig:S2}(a), atoms are dense in the dimer layers parallel to the $ab$ plane, while atoms are sparse between dimer layers. In BaCuSi$_2$O$_6$, which has a similar structure, the magnetic excitation is dispersionless along the $c$ direction, which indicates that the exchange interaction between neighboring dimer layers is negligible\,\cite{Sasago}. For these reasons, we infer that the exchange interaction between dimer layers is also negligible in Ba$_2$CoSi$_2$O$_6$Cl$_2$. Therefore, the present system can be described as an $S\,{=}\,1/2$ 2D coupled XXZ dimer model, as illustrated in Fig.\,\ref{fig:S2}(c).\\
\ \par

\section*{Effect of the spin-lattice coupling on the anisotropy of the exchange interaction}
From the analysis of the magnetization process, we found that the anisotropy of the interdimer exchange interaction, ${\alpha}^{\parallel}/{\alpha}^{\perp}\,{=}\,0.56$, is different from that of the intradimer exchange interaction, $J^{\parallel}/J^{\perp}\,{=}\,0.15$. In the first approximation, these anisotropies should be the same. In this section, we discuss the anisotropy of the exchange interaction between fictitious spins. 

The magnetic property of Co$^{2+}$ in an octahedral environment is determined by the lowest orbital triplet $^4T_1$.\cite{Abragam,Lines} This orbital triplet splits into six Kramers doublets owing to spin-orbit coupling and the uniaxial crystal field, which are expressed together as
\begin{eqnarray}
{\cal H}^{\prime}=-(3/2){\lambda}({\bm l}\cdot{\bm S})-{\delta}\left\{(l^z)^2-2/3\right\},
\label{eq:perturb}
\end{eqnarray}
where $\bm l$ is the effective angular momentum with $l=1$ and $\bm S$ is the true spin with $S=3/2$. When the temperature $T$ is much lower than the magnitude of the spin-orbit coupling constant ${\lambda}\,{=}\,{-}\,178$ cm$^{-1}$, i.e., $T\,{\ll}\, |{\lambda}|/k_{\rm B}\,{\simeq}\,250$\,K, the magnetic property is determined by the lowest Kramers doublet, which is given by $l^{\,z}\,{+}\,S^z\,{=}\,{\pm}1/2$, and the effective magnetic moment of Co$^{2+}$ is represented by ${\bm m}\,{=}\,g{\mu}_{\rm B}{\bm s}$ with the fictitious spin-1/2 operator $\bm s$~\cite{Abragam,Lines}. The true spin $\bm S$ is related to the fictitious spin $\bm s$ as
\begin{eqnarray}
S^x\rightarrow ps^x,\hspace{3mm} S^y\rightarrow ps^y,\hspace{3mm} S^z\rightarrow qs^z,
\label{eq:trans}
\end{eqnarray}
where $p$ and $q$ are real parameters determined by ${\lambda}$ and ${\delta}$.\cite{Lines} If the exchange interaction between true spins is described by the Heisenberg model ${\cal H}\,{=}\,J_{ij}({\bm S}_i\cdot{\bm S}_j)$, then the effective exchange interaction between fictitious spins is expressed as
\begin{eqnarray}
{\cal H}_{\rm eff}^{(0)}\hspace{-3mm}&=&\hspace{-3mm}J_{ij}\left[p^2\left\{s_i^xs_j^x+s_i^ys_j^y\right\}+q^2s_i^zs_j^z\right] \nonumber\\
\hspace{-3mm}&=&\hspace{-3mm}\left[J_{ij}^{\perp}\left\{s_i^xs_j^x+s_i^ys_j^y\right\}+J_{ij}^{\parallel}s_i^zs_j^z\right].
\label{eq:int0}
\end{eqnarray}
In Ba$_2$CoSi$_2$O$_6$Cl$_2$, the condition $p^2\,{>}\,q^2$ is realized, which leads to the XY like exchange interaction. Because the parameters $p$ and $q$ are determined by the spin-orbit coupling and the tetragonal crystal field, these parameters should be common to all the exchange interactions. Hence, additional effect is necessary to describe the difference between $J^{\parallel}/J^{\perp}$ and ${\alpha}^{\parallel}/{\alpha}^{\perp}$ in Ba$_2$CoSi$_2$O$_6$Cl$_2$. 

Here, we consider the local spin-lattice coupling that modulates the exchange interaction between true spins ${\bm S}$, as argued by Penc {\it et al.}\cite{Penc} to explain the one-half magnetization plateau observed in CdCr$_2$O$_4$.\cite{Ueda} The local spin-orbit coupling leads to the effective biquadratic exchange interaction of the form
\begin{eqnarray}
{\cal H}_{\rm bq}=-K_{ij}({\bm S}_i\cdot{\bm S}_j)^2,
\label{eq:int1}
\end{eqnarray}
with $K_{ij}\,{>}\,0$.\cite{Penc} Substituting the relations of eq. (2) into ${\cal H}_{\rm bq}$ of eq. (4), we obtain a quadratic exchange term
\begin{eqnarray}
{\cal H}_{\rm eff}^{(1)}=\frac{K_{ij}}{2}p^2\left[q^2\left\{s_i^xs_j^x+s_i^ys_j^y\right\}+p^2s_i^zs_j^z\right].
\label{eq:int2}
\end{eqnarray}
Here, we neglect the constant term. Comparing eqs. (3) and (5), we see that when the original exchange interaction ${\cal H}_{\rm eff}^{(0)}$ is of XY like, i.e. $p^2\,{>}\,q^2$, ${\cal H}_{\rm eff}^{(1)}$ becomes Ising like, and vice versa. Consequently, the total exchange interaction ${\cal H}_{\rm eff}\,{=}\,{\cal H}_{\rm eff}^{(0)}\,{+}\,{\cal H}_{\rm eff}^{(1)}$ goes toward isotropic form. We infer that there exists the local spin-lattice coupling that modulates the interdimer exchange interaction acting between true spins, so that the effective interdimer exchange interaction acting between fictitious spins becomes more isotropic than the effective intradimer exchange interaction.

\end{document}